\documentclass[twocolumn]{aastex631}

\usepackage{multirow}
\usepackage{amsmath}
\usepackage{amsbsy}
\usepackage{amsfonts}
\usepackage{amssymb}
\usepackage{amsthm}
\usepackage{yfonts}
\usepackage{enumerate}
\DeclareMathOperator{\cm}{cm}

\usepackage{xcolor}
\usepackage{soul}
\DeclareMathOperator{\m}{m}

\DeclareMathOperator{\N}{N}

\DeclareMathOperator{\J}{J}
\DeclareMathOperator{\s}{s}
\DeclareMathOperator{\g}{g}

\setstcolor{red}

\begin{document}

\title{Hot Jupiters are asynchronous rotators}

\author{Marek Wazny}
\affiliation{David A. Dunlap Department  of Astronomy \& Astrophysics, University of Toronto.
50 St.  George Street, Toronto, Ontario, M5S 3H4, Canada  }
\affiliation{Department of Physics, University of Toronto,
60 St George Street, Toronto, Ontario, M5S 1A7, Canada}
\author{Kristen Menou}

\affiliation{David A. Dunlap Department  of Astronomy \& Astrophysics, University of Toronto.
50 St.  George Street, Toronto, Ontario, M5S 3H4, Canada  
}
\affiliation{Department of Physics, University of Toronto,
60 St George Street, Toronto, Ontario, M5S 1A7, Canada}
\affiliation{Physics \& Astrophysics Group, Dept.   of  Physical  \&  Environmental  Sciences,  University  of  Toronto  Scarborough,\\   1265 Military Trail, Toronto, Ontario, M1C 1A4, Canada \\}
\affiliation{Observatoire de la Cote d'Azur, TOP team, Laboratoire Lagrange - CNRS, Nice, France}
\begin{abstract}
Hot Jupiters are typically assumed to be synchronously rotating, from tidal locking. Their thermally-driven atmospheric winds experience Lorentz drag on the planetary magnetic field anchored at depth. We find that the magnetic torque does not integrate to zero over the entire atmosphere. The resulting angular momentum feedback on the bulk interior can thus drive the planet away from synchronous rotation. Using a toy tidal-ohmic model and atmospheric GCM outputs for HD189733b, HD209458b and Kepler7b, we establish that off-synchronous rotation can be substantial at tidal-ohmic equilibrium for sufficiently hot and/or magnetized hot Jupiters. Potential consequences of asynchronous rotation for hot Jupiter phenomenology motivate follow-up work on the tidal-ohmic scenario with approaches that go beyond our toy model.

\end{abstract} 

\section{Introduction} \label{intro}
Hot Jupiters, a class of exoplanets with masses and radii comparable to Jupiter but with orbital periods less than 10 days, present a long-standing problem known as the 'radius anomaly’. A significant fraction of these planets possess radii in excess of $\sim 1.2$ Jupiter radii \citep{Charbonneau_2000,Henry_2000}, a phenomenon that cannot be interpreted by the standard theory of planetary evolution. This discrepancy between observed and theoretical radii is often referred to as the radius inflation problem. Several interpretations have been proposed to explain this anomaly, including tidal heating \citep{BODENHEIMER_2001_tide_dissapation,Bodenheimer_2003,Ibgui_2009,Ibgui_2010}, Ohmic heating \citep{Batygin_2010,2010showman,Perna_2010,Perna_2010b}, inward transfer of irradiation \citep{Tremblin_2017,Sainsbury-Martinez_2019}, and high opacity \citep{Burrows_2007}. Despite these proposals, the radius inflation problem for hot Jupiters remains one of the outstanding unresolved issues in our understanding of extrasolar planets \citep{Sarkis_2021}.

However, Ohmic heating has emerged as a promising solution to the radius inflation problem, with \citet{Thorngren_2018} suggesting that it is the most likely scenario. This mechanism, which involves the dissipation of electrical currents in the conductive interiors of these planets, has been successful in explaining the inflated radii of many hot Jupiters \citep{Wu_Lithwick,Thorngren_2021}. Still, \citet{Huang_2012} insist it may fall short when applied to the most inflated planets, indicating that additional processes may be at play.
The focus of the theory has largely been energetic aspects like Joule heating at depth to inflate the planets. However, there is also a need to consider the effects of drag and momentum balance in this context \citep{Rogers_2014,Rogers_2017, Beltz_2022}. In particular, the zonal winds in the atmosphere, which are driven by thermal gradients, experience drag as they interact with the planetary magnetic field. This means that any net momentum that is removed from the atmosphere must be deposited somewhere else, possibly deeper in the planet. 

In this work, we make the straightforward assumption that the net momentum that is removed from the atmosphere is deposited in the bulk interior. This would imply that the bulk of the planet can be torqued as a result of the magnetic drag acting in the atmosphere. This could lead to the possibility of asynchronous rotation for the bulk planetary interior, a scenario that has not been extensively explored in the context of hot Jupiters.

We investigate this tidal-ohmic scenario with a toy model,  combining tidal theory with simple momentum balance arguments. Our model relaxes the assumption of synchronous rotation and indeed suggests the possibility of significant asynchronous rotation, for sufficiently hot and magnetized hot Jupiters. The tidal-ohmic scenario opens up new {avenues} for understanding hot Jupiter observables by providing a channel for strong coupling between the interior and the atmosphere of these planets.

\section{Simulated Atmospheres}\label{data}

We employ PlaSim-Gen as a general circulation model (GCM) for tidally locked gas giants that simulates their atmospheric dynamics and was adapted for hot Jupiters by considering sufficiently hot atmospheres. This model solves meteorological fluid equations on atmospheric pressure levels to evolve temperature and wind velocity as 3-dimensional fields. These fields are solved for 1200 planet days until a near equilibrium is found, then we extract an average of the data over the last four days. For further model details see \citet{Paradise_2022, Menou2020}, and see \citet{Menou2022} for the specific data used throughout this work.
\begin{figure}
    \epsscale{1.2}
    \plotone{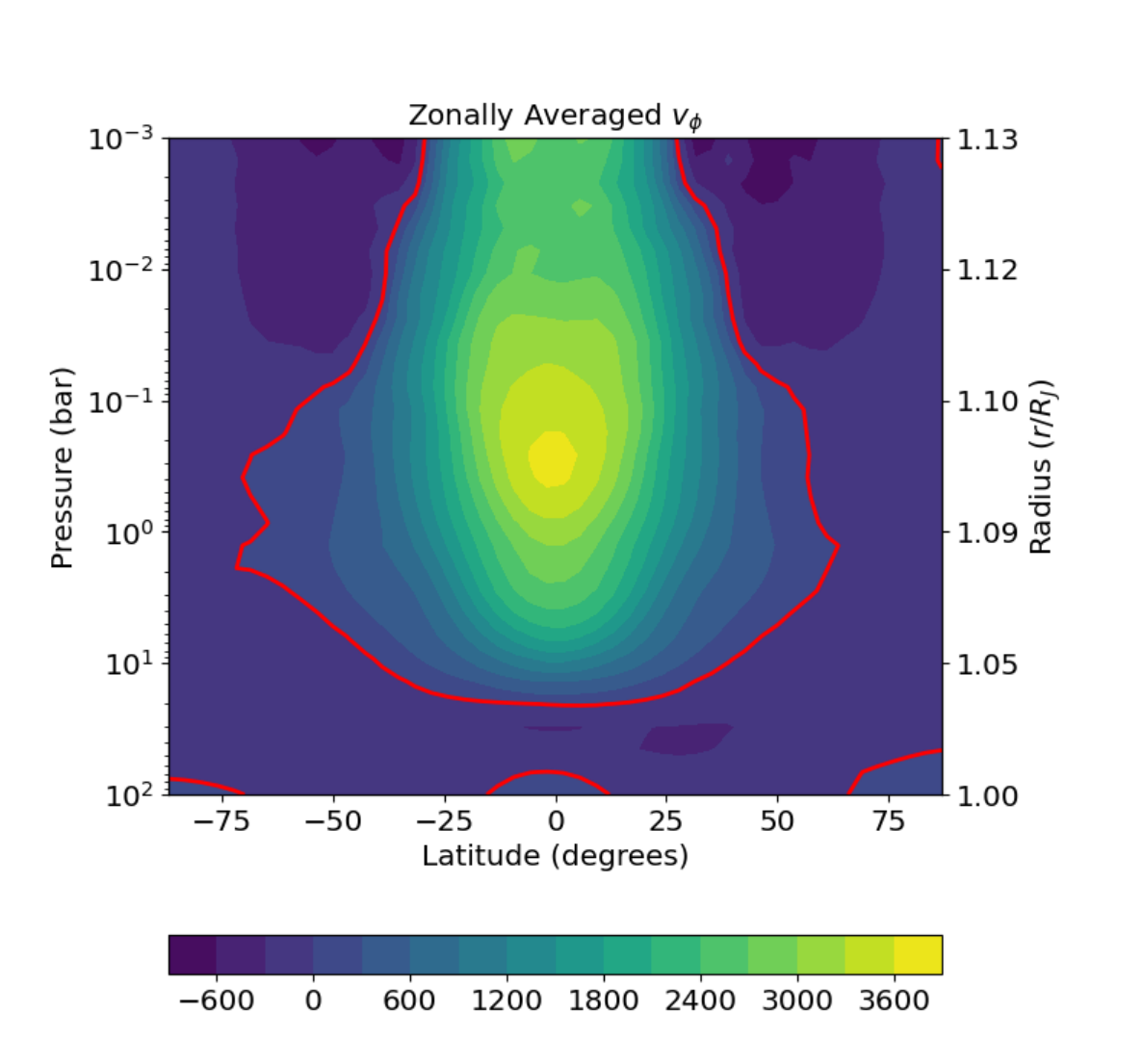}
    \caption{Zonally-averaged zonal wind speeds of planet Kepler7b generated by the GCM. Wind speeds are shown in m s$^{-1}$ where the red line separates regions of opposite wind directions. Wind speeds are solved on pressure levels seen on the left vertical axis and the converted radial coordinates (averaged) via Equation \ref{hydrostatic} are seen on the right vertical axis.}
    \label{winds}
\end{figure}

Simulated temperature and velocity fields from observed planets HD189733b, HD209458b, and Kepler7b are used to extrapolate other relevant properties for this analysis. Note that each field is evaluated as a function of pressure (1 mbar to 100 bar), latitude, and longitude. Where relevant, hydrostatic equilibrium is used to convert pressure, $p$ to radial coordinates, $r$. Substituting pressure for density via the ideal gas law gives
\begin{equation} \label{hydrostatic}
    \frac{dp}{dr} =  -g\frac{\mu m_{H}}{kT}p ,
\end{equation}
where $T$ is the temperature and assuming constant gravitational acceleration $g$, mean molecular weight $\mu$, Boltzmann constant $k$, and mass of hydrogen $m_H$. Although there is evidence to suggest that warmer hot Jupiters could be enriched in heavy elements \citep{Thorngren_2016,Welbanks_2019,Edwards_2023}, here for simplicity, we assume that atmospheric composition is approximately solar \citep{Hansen_2007} and adopt $\mu = 2.4$ throughout our analysis. Figure \ref{winds} shows the zonal wind in radius coordinates for planet Kepler7b.

The base radius for the deepest point of the atmosphere (100 bar) is fixed at 1 Jupiter radius. This assumption is crude but has a minor impact on our results relative to other factors such as magnetic field strength. We confirmed that modifying this base radius by up to 50\% does not fundamentally impact our general conclusions, but acknowledge that a more consistent treatment should be employed in future work. We assume a radiative-convective boundary located at 100 bar for simplicity, although this would need to be computed together with thermal state \citep{Fortney2021}.
\section{Magnetic Effects}
\subsection{Density, Ionization, and Resistivity}
To determine the density of the atmosphere we assume an ideal gas law. The atmospheric gas densities have mean values $1.12\times10^{-4}$ g cm$^{-3}$ for HD189733b and HD209458b while $1.22\times 10^{-4}$ g cm$^{-3}$ for Kepler7b. The variation in density is about 5 orders of magnitude with minimum values around $\sim 10^{-8}$ g cm$^{-3}$ in the upper atmosphere and maximum values reaching $\sim  10^{-3}$ g cm$^{-3}$ in the deeper atmosphere. This is typical of most hot Jupiter atmospheres. See also in \citet{Perna_2010}.

Hot Jupiter temperatures roughly range from 1000 $\sim$ 3000 K, hence thermal ionization of alkali metals provide the free electric charges. To estimate the ionization fraction, $x_e$, in a weakly ionized regime, the Saha equation is recast into a single ionization form \citep{Sato1991},
\begin{align}
    x_e \equiv \frac{n_e}{n_n} &= \sum_j \frac{n_j}{n}x_j ,\\
    \frac{x_j^2}{1-x_j^2} &\simeq \frac{1}{n_j kT}\left(\frac{2\pi m_e}{h^2}\right)^{3/2} (kT)^{5/2}\, \text{exp}\left(-\frac{I_j}{kT}\right) ,\nonumber
\end{align}
where $n_e$ and $n_n$ are the number densities of electrons and
neutrals, $n_j$ is the number density of element $j$, $I_j$ is the corresponding first ionization potential, $n$ is the total number density of the gas, and $m_e,h$ are the usual constants. We use the first 28 elements of the periodic table (H to Ni)
for the ionization potentials, assuming solar abundances. This solution for the ionization fraction allows the electric resistivity to be evaluated everywhere in the atmosphere as 
\begin{equation}\label{resistivity}
    \eta = 230 \frac{\sqrt{T}}{x_e}\,\cm^2\,\s^{-1} .
\end{equation}

The GCM outputs and Equation \ref{resistivity} provide the 3D resistivity field. The mean values are $2.76\times10^{29}$ cm$^2$ s$^{-1}$, $2.26\times10^{29}$ cm$^2$ s$^{-1}$ for HD189733b, HD209458b and $1.48\times 10^{27}$  cm$^2$ s$^{-1}$ for Kepler7b, with orders-of-magnitude spatial variations. Minimum values are similar for all planets $\sim 10^{12}$  cm$^2$ s$^{-1}$, while maximum values are $\sim 10^{33}$  cm$^2$ s$^{-1}$  for HD189733b and $\sim 10^{31} $  cm$^2$ s$^{-1}$ for Kepler7b and HD209458b.

\subsection{Induced Current and Magnetic Drag Time}\label{current}
It is believed that magnetic fields in hot Jupiters arise in a similar manner to that of Jupiter and Saturn, through an interior convection-driven dynamo. We assume the magnetic field geometry to be purely poloidal, in the form of an aligned dipole;
\begin{equation}\label{dipole}
    \mathbf{B}_{\text{dip}}(r,\theta) = B_0(2\cos\theta\,\mathbf{e}_r+\sin\theta\,\mathbf{e}_{\theta}) ,
\end{equation}
where $r$ is the radial coordinate, $\theta$ is the polar angle, and $B_0$ is the field strength. Surface planetary magnetic field strengths for hot Jupiters have been estimated to be anywhere from 1-100 G \citep{Cauley2019}. Since Ohmic dissipation scales as $B^2$, we choose a representative value of $B_0=10$ G in our analysis.

Charged zonal winds deform the poloidal magnetic field, generating a toroidal (east-west) component. The toroidal field obeys the resistive induction equation,
\begin{align}\label{induction}
    \frac{\partial B_{\phi}}{\partial t} &= r\sin\theta\left[\frac{\partial\Omega}{\partial r}B_r+\frac{1}{r}\frac{\partial\Omega}{\partial\theta}B_{\theta}\right] +\frac{1}{r}\frac{\partial}{\partial r}\left[\eta \frac{\partial}{\partial r}(rB_{\phi})\right]\nonumber
    \\&\hspace{2mm}+\frac{1}{r^2}\frac{\partial}{\partial \theta}\left[\frac{\eta}{\sin\theta}\frac{\partial}{\partial\theta}(\sin\theta B_{\phi})\right] ,
\end{align}
where $\Omega =\text{v}_{\phi} (r\sin\theta)^{-1}$ and $\text{v}_{\phi}$ is the zonal wind speed. Because the entire velocity field, $\mathbf{v}$, is dominated by v$_{\phi}$ from largely zonal jets, we narrow our analysis by neglecting gradient polodial terms $\partial/\partial\theta$ in Equation (\ref{induction}). This is consistent with neglecting the magnetic drag applied to the meridional (north-south) winds, which are small.  

An alternative view of the MHD problem is that an induced current $\mathbf{j}$, is supporting the induced toroidal field. The benefits of this approach will become clear below. Neglecting polodial terms in the induction equation is equivalent to $\mathbf{j}_{\theta}\gg \mathbf{j}_r$. Imposing steady state to Equation (\ref{induction}) leads to the induced current \citep{LIU2008653},  

\begin{align}\label{jtheta}
    j_{\theta}(r,\theta,\phi) &= -\frac{c\sin\theta}{4\pi r \eta}\int_r^R dr'r'^2\left(\frac{\partial\Omega}{\partial r'}B_r+\frac{1}{r'}\frac{\partial\Omega}{\partial\theta}B_{\theta}\right) \nonumber
    \\&\hspace{5mm}+\frac{R}{r\eta}\big[\eta j_{\theta}\big]_{r = R} ,
\end{align}
where $c$ is the speed of light. The resistivity, $\eta(r,\theta,\phi)$ depends on position through temperature, and $ \big[\eta j_{\theta}\big]_{r = R}$ is an outer boundary condition related to the ionosphere. 
\begin{figure}
    \epsscale{1.15}
    \plotone{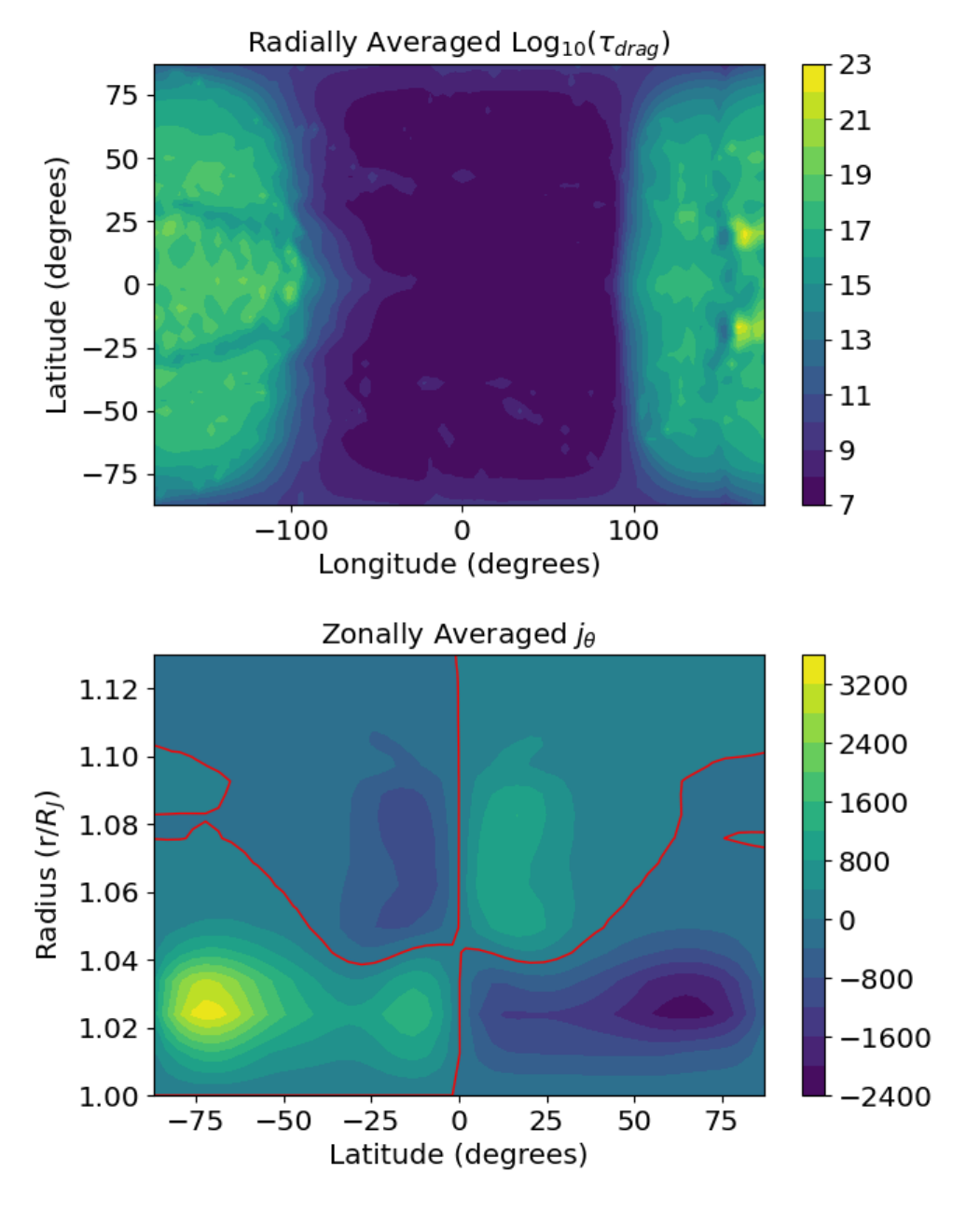}
    \caption{Top: Numerically evaluated Log$_{10}$ drag timescale (s) averaged radially. Bottom: Numerically integrated current in G s$^{-1}$ using (Equation \ref{jtheta}) with zero boundary term and red lines separating positive and negative regions.} Both plots are for the planet Kepler7b, assuming a magnetic field strength of 10~G.
    \label{jthetazon}
\end{figure}

\begin{figure*}
    \epsscale{1.15}
    \plotone{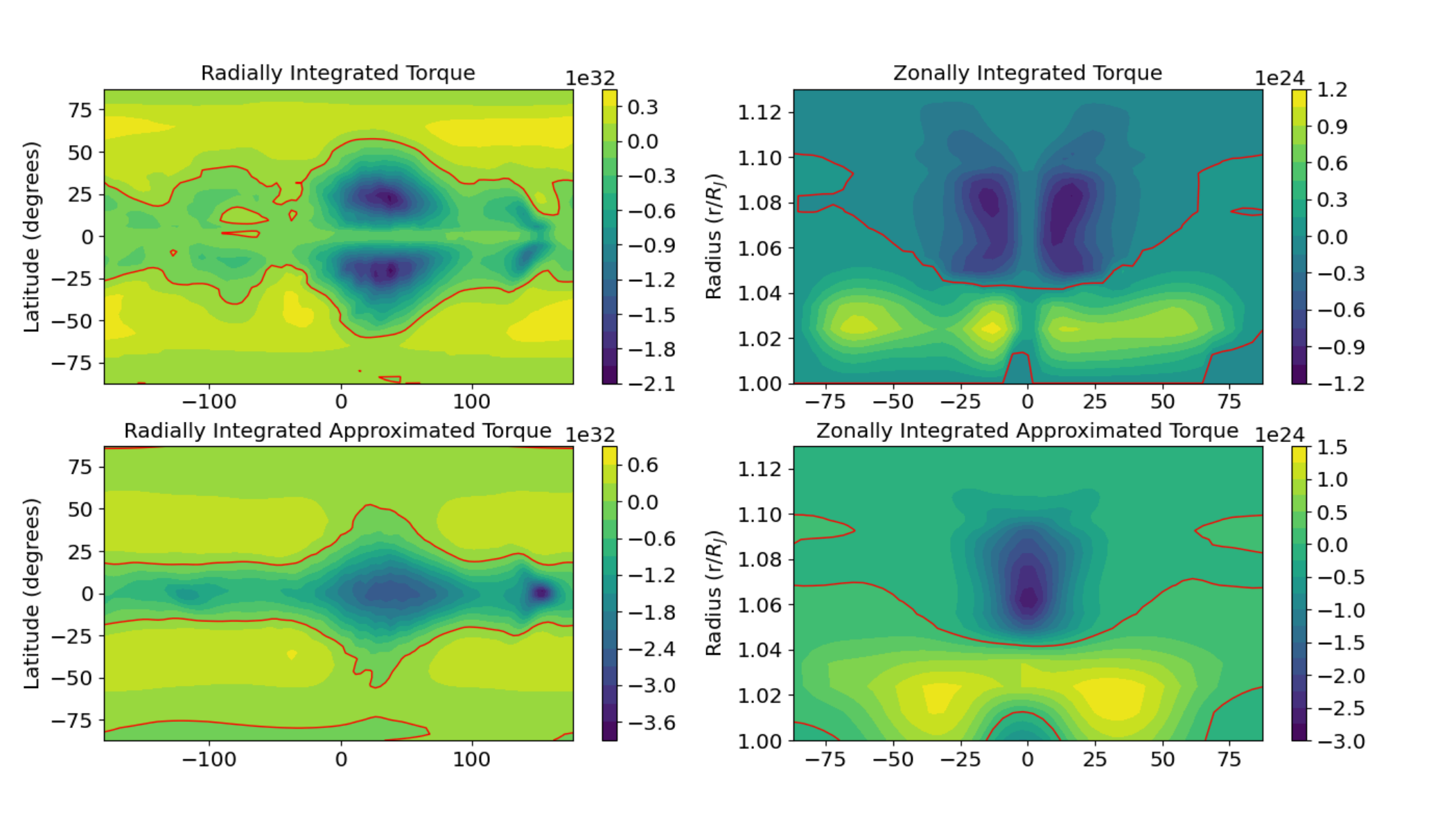}
    \caption{Left: Magnetic torque (dyne cm) of planet Kepler7b radially evaluating Equation \ref{gamma} for the approximate drag time in Equation \ref{tau approx} (bottom) and Equation \ref{tau} (top). Right: Magnetic torque of planet Kepler7b zonally integrated for the approximate (bottom) and integrated (top) drag times. Red lines separate regions of positive and negative contributions.}
    \label{torque}
\end{figure*}

We will ignore this last term. Although this is a valid assumption in neutral-top atmospheres (such as those of Jupiter and Saturn) with a top insulating boundary, in the ionized atmospheres of hot Jupiters this assumption may not hold. We address this by simply noting that any additional contribution from this term may increase the magnetic drag and hence the total torque. As our work is concerned with setting up a minimum threshold for asynchronous rotation, neglecting this term is a reasonable working assumption.

Previous works \citep{Perna_2010,Menou_2012}, evaluate this induced current through an order of magnitude approximation: $j_{\theta} \sim c\text{v}_{\phi} B_0/(4\pi\eta)$. Here, we numerically integrate Equation (\ref{jtheta}) without the boundary term, using the data in Section \ref{data}. The bottom of Figure \ref{jthetazon} shows the resulting induced current averaged across longitude for planet Kepler7b. 

\begin{nolinenumbers}
\begin{table*}[t]
    \centering
    \caption{Below are the mean planet characteristics and magnetic properties for planets HD189733b, HD209458b, and  Kepler7b. The magnetic field is assumed to have a dipole geometry with a strength of 10 Gauss for each planet.}
\begin{tabular}{ p{2cm} p{2cm} p{2cm} p{2cm} p{5.5cm}}
 \hline 
Parameter & HD189733b & HD209458b & Kepler7b & Description\\ 
 \hline
 $g$  &2200& 800& 414 & Gravitational acceleration (cm s$^{-1}$)\\
  $T_{eq}$ & 1200 & 1450& 1630 & Equilibrium temperature (K)\\
$\overline{\rho}$ & 1.18$\times 10^{-3}$& 1.17 $\times 10^{-3}$& 1.22 $\times 10^{-3}$  & Mean density  (g cm$^{-3}$)\\
$\overline{\eta}$ & 2.76 $\times 10^{29}$ & 2.26 $\times 10^{27}$& 1.48 $\times 10^{27}$  & Mean resistivity (cm$^{2}$ s$^{-1}$)\\ 

$\overline{\tau}_{\text{drag}}$   & 3.24 $\times 10^{22} $ &  $ 5.02 \times 10^{19}$&  1.41$\times 10^{19}$& Mean drag time (s) \\
$\Gamma_{\text{tot}}$ & $-5.89\times 10^{31}$ &  $-5.19\times 10^{31}$&  $-3.58\times 10^{32}$ & Total torque (dyne cm)\\
 $\Delta\omega/\omega_\text{sync}$ &  $0.72$ &  $2.6$&  $ 31$ & Asynchronicity ratio \\
 \hline
\end{tabular}
    \label{table}
\end{table*}
\end{nolinenumbers}

With the induced current, we can analyze the magnetic field effects on zonal winds. In general, calculating this effect requires solving MHD equations, for example as in \citet{Rogers_2014}. However, our approach estimates MHD effects through the timescale, $\tau_{\text{drag}}$. This timescale evaluates the effective time for the zonal winds to reach stationarity against the toroidal magnetic field.

A magnitude estimate for the drag timescale can be found by adopting an ion drag formulation, as outlined in \citet{2005AnGeo..23.3313Zhu}. Using the estimate that the momentum carried by the wind is proportional to the Lorentz force, 
while assuming $j_{\theta} \sim c\text{v}_{\phi} B_0/(4\pi\eta)$, leads to the order of magnitude approximation,
\begin{equation}\label{tau approx}
    \tau_{\text{drag}} \sim \frac{4\pi\rho\eta}{B_0^2\cos\theta} .
\end{equation}
Using Equation (\ref{tau approx}), magnetic drag has been found to be significant in the atmospheres of hot Jupiters \citep{Perna_2010}.
This work expands on the approximation by using the induced current obtained in Equation (\ref{jtheta}) to calculate the drag time more consistently through
\begin{equation}\label{tau}
    \tau_{\text{drag}} \sim \frac{\rho \mid\text{v}_{\phi}\mid c}{\mid\mathbf{j}_{\theta}\times \mathbf{B}\mid} .
\end{equation} 

Using this new prescription for the drag time, results still generally agree with those reported in \citet{Perna_2010}, with a few notable differences. This includes longer drag times deeper in the atmosphere and on the night-side, as well as shorter drag times on the day-side. The top of Figure \ref{jthetazon} shows the drag time averaged radially for planet Kepler7b.

\subsection{Magnetic Torque}\label{mag}
The magnetic drag timescale approximates the drag experienced by the winds as a linear term which takes the form $\text{v}_{\phi}/\tau_{\text{drag}}$.

Drag forces induce a torque on the atmosphere which determines the momentum transfer from the MHD effects on the wind. Using the linear drag term, the total magnetic torque can be computed as
\begin{equation}\label{gamma}
    \Gamma_{\text{tot}}  = -\int d^3\mathbf{r}\,\rho (\mathbf{r})\mathbf{r}\,\frac{\text{v}_{\phi}(\mathbf{r})}{\tau_{\text{drag}}(\mathbf{r})} .
\end{equation}
In Equation (\ref{gamma}) integration is done over the atmosphere, rather than the entire planet since no wind is present in the bulk interior, by design. Our sign convention highlights that the total torque works against the dominant winds.

Figure \ref{torque} visualizes the magnitude of the torque field in Equation (\ref{tau}). The greatest contributions occur where zonal jets and higher temperatures are present. The right side of Figure \ref{torque} shows that the geometry of the induced current in Figure \ref{jthetazon} largely determines the geometry of the zonal torque.

\section{Tidal-Ohmic Scenario}\label{sec4}

We posit that the total torque experienced by the atmospheric winds exerts a continuous reciprocal torque on the bulk rotation rate of the planet. This assumes no external loss of angular momentum, such as in an outflowing MHD wind.  Conservation of angular momentum across the atmosphere and interior regions implies deposition of angular momentum (via a positive torque contribution) in the core of the planet, where it would presumably be well mixed by convective motions. While we do not detail a specific mechanism for this momentum transfer, perturbed magnetic field lines rooted in the planet’s interior dynamo region are prime candidates {\citep[see, e.g., Appendix in][]{Wu_Lithwick}} .

Conservation of angular momentum in the atmosphere yields $d(I_{\text{a}}\omega_{\text{a}}) /dt = -\Gamma_{\text{tot}}$,
where $I_{\text{a}}$ is the moment of inertia and $\omega_{\text{a}}$ is the angular frequency of the atmosphere. For simplicity, we assume that the planet does not deform, hence $I_{\text{a}}$ is constant.

For the core dynamics, with core mass $M_{\text{c}}$ and radius $R_{\text{c}}$, we adopt the time-independent moment of inertia for Jupiter, $I_{\text{c}} \sim 0.264M_cR_c^2$ \citep{HELLED_2011}. The synchronous angular frequency, $\omega_{\text{sync}}$, is equal to the orbital frequency, so that the degree of asynchronicity (off-synchronous frequency) can be written as $\Delta\omega = \omega_{\text{c}}-\omega_{\text{sync}}$.  We neglect any tides in the atmosphere and assume that tides only act on the angular frequency of the core, $\omega_{\text{c}}$. By analogy with magnetic drag, we introduce a tidal timescale, $\tau_{\text{tide}}$, which represents the typical timescale on which the core’s angular frequency is synchronized by tidal effects. Following \citet{2010ApJ...723..285Hansen1} and \citet{Hansen_2012}, for an orbital period $P$, and a tidal dissipation parameter $Q$, the tidal timescale can be written as $\tau_{\text{tide}} \sim QP$ where
\begin{equation}
    Q = 3.0\times 10^8 \left( \frac{a}{0.1\,\text{AU}}\right)^{3/2}\left(\frac{R_p}{1\,R_{J}} \right)^{-5}\left(\frac{M_*}{M_{\odot}}\right)^{-1/2} .
\end{equation}
Here $a$ is the semi-major axis, $R_p$ is the radius of the planet, and $M_*$ is the mass of the host star. We acknowledge that adopting other tidal models \citep{GOLDREICH1966375,Leconte2010} would impact our results, although likely not our main conclusions.

Given this tidal timescale, conservation of angular momentum for the core yields
\begin{equation}\label{momentum}
    I_{\text{c}}\frac{d\omega_{\text{c}}}{dt} = \Gamma_{\text{tot}}+0.264M_{\text{c}}R_{\text{c}}^2\frac{\Delta\omega}{\tau_{\text{tide}}} ,
\end{equation}
under the combined action of tidal and magnetic torques.
At steady state, $d\omega_{\text{c}}/dt=0$, so that the off-synchronous frequency can be evaluated from the total torque entering Equation (\ref{momentum}) as 
\begin{equation}
    \Delta \omega = -\frac{1}{0.264}\frac{ \tau_{\text{tide}}}{M_{\text{c}}R_{\text{c}}^2} \Gamma_{\text{tot}} .
\end{equation}
 When the off-synchronous angular frequency, $\Delta\omega$, approaches or exceeds the magnitude of the orbital frequency, $\omega_{\text{sync}}$, tidal locking is significantly disrupted. We address this possibility for the planets Kepler-7b, HD189733b, and HD209458b, with estimates of $\Delta\omega/\omega_\text{sync}$ for each planet reported in Table \ref{table}.

\begin{figure*}
    \epsscale{1.15}
    \plotone{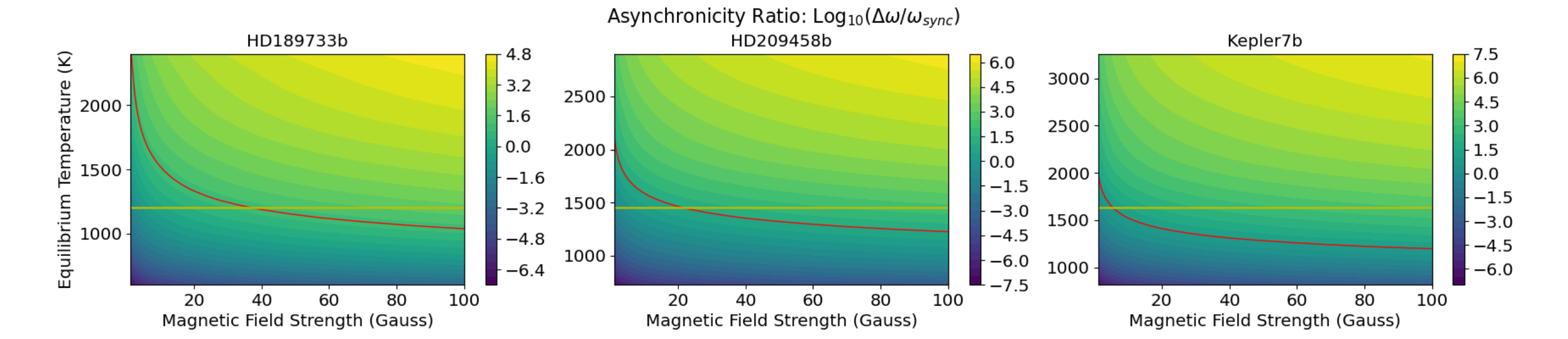}
    \caption{Asynchronicity ratio, Log$_{10}(\Delta \omega/\omega_{\text{sync}})$, for planets HD189733b, HD209458b, and Kepler7b. Values are calculated given an equilibrium temperature and magnetic field strength. Red lines correspond to the isolines $\Delta \omega/\omega_{\text{sync}}=1$ while yellow lines show the equilibrium temperature for each planet.}
    \label{omega}
\end{figure*}

We take parameters $M_* = 0.83\, M_\odot$, $a = 0.0313$ au, $P = 1.9 \times 10^5$ s for HD189733b, $M_* = 1.07 \,M_\odot$, $a = 0.0463$ au, $P = 3.04 \times 10^5$ s for HD209458b \citep{Rosenthal_2021}, and $M_* = 1.1\, M_\odot $, $a = 0.0624$ au, $P = 4.2 \times 10^5$ s for Kepler7b \citep{Latham_2010,Stassun_2019}. Consider the case of Kepler7b with an orbital frequency of $ 1.5 \times 10^{-5}$ s$^{-1}$. This provides a large off-synchronous ratio as seen in Table \ref{table}, indicating potentially strong asynchronicity for this specific hot Jupiter. By contrast, HD189733b has an orbital frequency of $3.3\times 10^{-5}$ s$^{-1}$, giving an off-synchronous ratio less than one (Table \ref{table}). This amounts to a small, but non-negligible amount of asynchronicity, even for this cooler planet. Finally, planet HD209458b is a borderline case with an orbital frequency of $2.1\times 10^{-5}$ s$^{-1}$ leading to an off-synchronous ratio $\sim 1$ (Table \ref{table}).

Equilibrium temperatures near or above 1500 K appear to be needed for strong tidal-ohmic effects, unless the planetary magnetic field strengths are well in excess of the $10$~G assumed here. To illustrate this, we calculate the asynchronicity ratio for a broad range of temperatures and magnetic field strengths given each planet. This results in Figure \mbox{\ref{omega}}, demonstrating the conditions needed for strong asynchronous rotation.

It is important to acknowledge that our results do not account for the coupling between winds and Lorentz torque. This issue has been investigated by various authors, yielding varying conclusions \citep{RogersKomeck2014,Rogers_2017, Koll_2018, Knierim2022,Beltz_2022}. In the context of the present work, although the details of this coupling are not fully understood, we can expect wind speeds to be reduced in the high-temperature limit, potentially leading to a reduction in the net torquing of the bulk planetary interior. Consequently, our results should be conservatively interpreted as providing an upper limit on the degree of asynchronous rotation that would emerge from a more complete, fully MHD formulation of the problem—especially at high temperatures, where magnetic drag is expected to have a strong influence on wind patterns in hot Jupiters.

\section{Conclusion}\label{summary}

Our work challenges the assumption of synchronous rotation for hot Jupiters in the presence of atmospheric Lorentz drag. The net angular momentum torque mediated by the planetary field lines does not cancel out over the atmosphere, suggesting that the reciprocal torque could drive the bulk interior away from synchronous rotation. We find that this off-synchronous rotation can be significant, particularly in hot Jupiters that are sufficiently hot and/or have strong magnetic fields. These findings are supported by a tidal-ohmic toy model, in combination with atmospheric GCM outputs for planets HD189733b, HD209458b, and Kepler7b, showing that the degree of asynchronicity is notable at tidal-ohmic equilibrium.

This work highlights the need to re-evaluate the rotational states of hot Jupiters and consider the broad implications of asynchronous rotation on observable properties, such as atmospheric features and phase curves \citep{Rauscher_Kempton_2014}. Our results motivate further research with more comprehensive models of the tidal-ohmic scenario, such as iterating magnetic drag effects into the GCM, which could yield deeper insights into the physical processes governing hot Jupiters, including the longstanding radius inflation problem. \\

\section*{Acknowledgements}
KM is supported by the National Science and Engineering Research Council of Canada. KM thanks C. Gammie for sharing an early implementation of the Saha ionization balance solver used in this work. 

We would furthermore like to acknowledge that this work was performed on land which for thousands of years has been the traditional land of the Huron-Wendat, the Seneca and the Mississaugas of the Credit. Today this meeting place is still the home to many indigenous people from across Turtle Island and we are grateful to have the opportunity to work on this land.

\bibliographystyle{aasjournal}
\bibliography{References}

\end{document}